\documentclass[A4paper, 11pt]{article}
\usepackage{amsfonts}
\usepackage{dsfont}
\usepackage{mathrsfs}
\usepackage{amssymb}
\usepackage{bbm}
\usepackage{epsfig}
\usepackage{amsmath}
\usepackage{appendix}
\usepackage{float}
\usepackage{color}
\usepackage[english]{babel}

\def\Reals{\mathop{\hbox{\mit I\kern-.2em R}}\nolimits}

\def\Complexes{{\hbox{\mit C\kern-.46em
            \vrule depth 0ex height 1.4ex width .05em\kern.41em}}}

\newtheorem{thm}{Theorem}
\newtheorem{defn}{Definition}
\newtheorem{coro}{Corollary}
\newtheorem{lem}{Lemma}
\newtheorem{remark}{Remark}

\title{\bf Finite-time and Asymptotic Convergence of \\ Distributed Averaging and Maximizing  Algorithms\footnote{This work has been supported in part
by the Knut and Alice Wallenberg Foundation, the Swedish Research
Council and  KTH SRA TNG.}}
\date{}

\author{Guodong Shi and Karl Henrik Johansson\thanks{The authors are with ACCESS Linnaeus Centre, School of Electrical Engineering,
Royal Institute of Technology, Stockholm 10044, Sweden.
       Email: {\tt\small $\{$guodongs,  kallej$\}$@kth.se}}}

\begin{document}

\maketitle
\begin{abstract}
In this paper, we formulate and  investigate a generalized consensus algorithm which makes an attempt to unify     distributed averaging and maximizing   algorithms considered in the literature. Each node iteratively updates its  state as a time-varying weighted average of its own state,  the minimal state,  and the maximal  state  of its neighbors. We prove that finite-time consensus is  almost impossible for averaging under this uniform model. Both time-dependent and state-dependent graphs are considered, and various necessary and/or sufficient conditions are presented on the consensus convergence. For time-dependent graphs, we show that quasi-strong connectivity is critical for  averaging, as is strong connectivity for  maximizing. For state-dependent graphs defined by a $\mu$-nearest-neighbor rule, where each node interacts with  its $\mu$ nearest smaller neighbors  and the  $\mu$ nearest larger neighbors, we show that  $\mu+1$ is a critical threshold on the total  number of  nodes for the  transit from finite-time to asymptotic convergence for averaging, in the absence of  node self-confidence. The threshold is $2\mu$ if each node chooses to connect only to neighbors with unique values. Numerical examples illustrate the tightness of the
conditions.  The results characterize some  fundamental similarities and differences between distributed averaging and maximizing   algorithms.
\end{abstract}


\section{Introduction}
Distributed averaging algorithms, where each node iteratively  averages its neighbors' states,   have been extensively studied in the literature, due to its wide applicability in  engineering \cite{tsi,jad03,cortes}, computer science \cite{cs2,cs3}, and social science \cite{degroot, social1,golub}. Recently also the  max-consensus algorithms have  attracted  attention. These algorithms compute the maximal value among the nodes, and have been used for leader election, network size estimation, and various applications  in wireless networks \cite{cortes,julien2}.

 The convergence to a consensus is central  in the study of averaging and maximizing algorithms but can be hard to analyze, especially when the node interactions are carried out over a switching graph. The most convenient  way of modeling the switching node interactions is just to assume the communication graphs are defined by a sequence of time-dependent graphs over the node set. The connectivity of this sequence of graphs plays an important role for the network to reach consensus. Joint connectivity, i.e.,  connectivity of the union graph over time intervals,  has been considered,  and various  convergence  conditions  have been established \cite{tsi, vb2, jad03, saber04, ren, mor,caoming1,caoming2,shi09,mor}. The switching topology  can be dependent on the  node states. For instance, in Krause's model, each node is connected only to  nodes within a certain distance \cite{krause}. Vicsec's model   has a similar setting but with  higher-order node dynamics \cite{vic95}. Because the node dynamics is coupled with the graph dynamics for state-dependent graphs, the  convergence analysis  is quite challenging.  Deterministic consensus algorithms with state-dependent graph were studied in \cite{vb1,julien}, and convergence results under  probabilistic models were established in \cite{tang,liuguo}.

Few studies  have discussed the  fundamental similarities and differences  between  distributed averaging and maximizing. Averaging and maximizing consensus algorithms are both  distributed  information processing over graphs, where nodes communicate and exchange information with its neighbors in the aim of collective convergence. Average consensus algorithms in the literature are based on two standing assumptions: local cohesion and node self-confidence.  The node states iteratively update as a weighted average of its neighbors' states, with a positive lower bound for  the weight corresponding to its own state \cite{tsi,vb2,jad03,ren,caoming1,caoming2,tsi11}.  Average consensus algorithms can also be viewed as the equivalent state evolution process where  each node updates its  state as a weighted average of its own state, and  the minimum  and maximum  states  of its neighbors. Maximizing (or minimizing) consensus algorithms  are simply based on that each node  updates its state to  the maximal (minimal) state among its neighbors \cite{max1,max2}. Asymptotic convergence is common in the study of averaging consensus algorithms \cite{caoming1,ren,tsi11,jad03}, while it has been shown that maximizing algorithms converge in general  in finite time \cite{max1,max2}. Finite-time convergence of averaging  algorithms was investigated in \cite{cortes} for a   continuous-time model, and recently finite-time consensus in discrete time was discussed  in \cite{it} for a special case of  gossiping \cite{boyd}.

 In this paper, we make the simple observation that averaging and maximizing algorithms can be viewed as instances of a more general distributed processing model. Using this model the transition of the consensus convergence can be studied for the two classes of distributed algorithms in a unified way. Each node iteratively updates its  state as a weighted average of its own state together with the minimum and maximum  states  of its neighbors. By special cases for the weight parameters, averaging and maximizing algorithms can be analyzed.
 Under this uniform model, we prove for averaging that finite-time consensus is  impossible in general, and asymptotical consensus   is possible only when the node self-confidence satisfies a divergence condition. Both time-dependent  and state-dependent graphs are considered, and various necessary and/or sufficient conditions are presented on the consensus convergence. For time-dependent graphs, we show that quasi-strong connectivity (existence of a spanning tree) is critical for  averaging, as is strong connectivity for  maximizing.   We use a $\mu$-nearest-neighbor rule to generate  state-dependent graphs, in which each node interacts with  its $\mu$ nearest smaller neighbors ($\mu$ neighbors with smaller state values), and the nearest $\mu$ larger neighbors. This model is motivated from recent studies of collective bird behavior \cite{pnas}. For averaging algorithms without node self-confidence under such state-dependent graphs, we show that $\mu+1$ is a critical value for the total number of  nodes: finite-time consensus is achieved globally if the number of nodes is no larger  than $\mu+1$, and finite-time consensus fails for almost all initial conditions if the number of nodes is larger than $\mu+1$. Moreover, it is shown that this critical number of nodes is instead  $2\mu$ if each node chooses to connect only to neighbors with distinct  values in the  neighbor rule.

  The rest of the paper is organized as follows. In Section 2 we introduce the considered network model, the uniform processing algorithm, and the consensus problem. The impossibilities of finite-time or asymptotic consensus are studied in Section 3. Consensus convergence  is studied for  time-dependent and state-dependent  graphs in Sections 4 and 5, respectively. We show numerical examples in Section 6 and finally some concluding remarks are given in Section 7.

\section{Problem Definition}
In this section, we introduce the network model, the considered algorithm, and define the problem of interest.

\subsection{Network}
We first recall some concepts and notations in graph theory \cite{god}. A directed graph (digraph) $\mathcal
{G}=(\mathcal {V}, \mathcal {E})$ consists of a finite set
$\mathcal{V}$ of nodes and an arc set
$\mathcal {E}\subseteq \mathcal{V}\times\mathcal{V}$.  An element $e=(i,j)\in\mathcal {E}$ is called an
{\it arc}  from node $i\in \mathcal{V}$  to $j\in\mathcal{V}$. If the
arcs are pairwise distinct in an alternating sequence
$ v_{0}e_{1}v_1e_{2}v_{2}\dots e_{k}v_{k}$ of nodes $v_{i}\in\mathcal{V}$ and
arcs $e_{i}=(v_{i-1},v_{i})\in\mathcal {E}$ for $i=1,2,\dots,k$,
the sequence  is called a (directed) {\it  path} with {\it length} $k$.  If there exists a path from node $i$ to node $j$,
then node $j$ is said to be reachable from node $i$.  Each node is thought to be reachable by itself. A node $v$ from which any other node is
reachable  is called a {\it center} (or a {\it root}) of $\mathcal {G}$. A digraph $\mathcal
{G}$ is said to be {\it strongly connected} if node $i$ is reachable from  $j$ for any two nodes $i,j\in\mathcal{V}$; {\it
quasi-strongly connected} if $\mathcal {G}$ has a center
\cite{ber}. The {\it distance} from $i$ to $j$ in a digraph $\mathcal{G}$, $d(i,j)$,   is the length of a shortest simple path $i \rightarrow j$ if $j$ is reachable from $i$, and the {\it diameter} of $\mathcal
{G}$ is $\rm{diam}(\mathcal{G})$$=\max\{d(i,j)|i,j \in\mathcal
{V},\ j\mbox{ is reachable from}\ i\}$. The
union of two digraphs with the same node set $\mathcal {G}_1=(\mathcal {V},\mathcal {E}_1)$ and $\mathcal
{G}_2=(\mathcal {V},\mathcal {E}_2)$ is defined as $\mathcal {G}_1\cup\mathcal
{G}_2=(\mathcal {V},\mathcal {E}_1\cup\mathcal {E}_2)$. A digraph $\mathcal {G}$ is said to be bidirectional if for every two nodes $i$ and $j$, $(i,j)\in \mathcal{E}$  if and only if $(j,i)\in \mathcal{E}$ .  A bidirectional graph $\mathcal{G}$ is said to be {\it connected} if there is a path between any two nodes.

Consider a network with node set $\mathcal{V}=\{1,2,\dots,n\}$, $n\geq 3$. Time is slotted. Denote the state of node $i$ at time $k\geq 0$ as $x_i(k)\in\mathds{R}$. Then $x(k)=\big(x_1(k) \dots x_n(k)\big)^T$ represents the network state.  For time-varying graphs, we use the following definition.
\begin{defn}[Time-dependent Graph] The interactions among the nodes are determined by a given sequence of digraphs with node set $\mathcal{V}$, denoted as $\mathcal{G}_k=
 (\mathcal{V}, \mathcal{E}_k)$, $k=0,1,\dots$.
\end{defn}

Throughout this paper, we call node $j$ a {\em neighbor} of node $i$ if there is an arc from $j$ to $i$ in the graph. Each node is supposed to always be a neighbor of itself.
Let $\mathcal{N}_i(k)$ represent the neighbor set of node $i$ at time $k$.

 \subsection{Algorithm}

The classical  average consensus algorithm in the literature is given by
\begin{align}\label{r1}
x_i(k+1)=\sum_{j\in\mathcal{N}_i(k)}a_{ij}(k)x_j(k),\ \ i=1\dots,n.
\end{align}
Two  standing assumptions  are fundamental in determining the nature of its dynamics:
\begin{itemize}
\item[{\bf A1}] {\it (Local Cohesion)} $\sum_{j\in\mathcal{N}_i(k)}a_{ij}(k)=1$  for all $i$ and $k$;
\item[{\bf A2}] {\it (Self-confidence)} There exists a constant $\eta>0$ such that $a_{ii}(k)\geq \eta$  for all $i$ and $k$.
\end{itemize}

These assumptions  are widely imposed  in the existing works, e.g., \cite{jad03,tsi,tsi2,tsi11,caoming1,caoming2,ren,vb2}. With A1 and A2, we have
  \begin{align}\label{102}
\sum_{j\in\mathcal{N}_i(k)}a_{ij}(k)x_j(k)=\eta x_i(k) + \big(a_{ii}(k)-\eta\big)x_i(k) +\sum_{j\in\mathcal{N}_i(k), j\neq i}a_{ij}(k)x_j(k)
  \end{align}
and
  \begin{align}\label{100}
\big(1-\eta\big)\min_{j\in \mathcal{N}_i(k)}x_j(k)\leq  \big(a_{ii}(k)-\eta\big)x_i(k) +\sum_{j\in\mathcal{N}_i(k), j\neq i}a_{ij}(k)x_j(k)\leq \big(1-\eta\big)\max_{j\in \mathcal{N}_i(k)}x_j(k).
  \end{align}

Noting the fact that for any $c\in[a,b]$ there exists a unique $\lambda\in[0,1]$ satisfying  $c=\lambda a+(1-\lambda)b$, we see from (\ref{100}) that for every  $i$ and $k$, there exists $\beta_k^{\langle i\rangle}\in [0,1]$ such that
 \begin{align}\label{101}
 &\big(a_{ii}(k)-\eta\big)x_i(k) +\sum_{j\in\mathcal{N}_i(k), j\neq i}a_{ij}(k)x_j(k)\nonumber\\
 &= \beta_k^{\langle i\rangle}\big(1-\eta\big)\min_{j\in \mathcal{N}_i(k)}x_j(k)+   \big(1-\beta_k^{\langle i\rangle}\big)\big(1-\eta\big)\max_{j\in \mathcal{N}_i(k)}x_j(k)\nonumber\\
 &= \alpha^{\langle i\rangle}_k\min_{j\in \mathcal{N}_i(k)}x_j(k)+\big(1-\eta-\alpha^{\langle i\rangle}_k\big)\max_{j\in \mathcal{N}_i(k)} x_j(k),
  \end{align}
where $\alpha^{\langle i\rangle}_k=\beta_k^{\langle i\rangle}(1-\eta)\in[0,1-\eta]$.

 Therefore, in light of (\ref{102}) and (\ref{101}),  based on  assumptions A1 and  A2,  we can always write the average consensus algorithm (\ref{r1})  into the following equivalent form:
 \begin{align}\label{r2}
x_i(k+1)=\eta x_i(k)+\alpha^{\langle i\rangle}_k\min_{j\in \mathcal{N}_i(k)}x_j(k)+\big(1-\eta-\alpha^{\langle i\rangle}_k\big)\max_{j\in \mathcal{N}_i(k)} x_j(k),
\end{align}
where $\alpha_k^{\langle i\rangle}\in[0,1-\eta]$ for all $i$ and $k$. Thus,  the information processing principle behind distributed averaging is  that each node iteratively takes a weighted average of its current state and the minimum   and  maximum  states of its neighbor set.

The standard maximizing algorithm \cite{max1, max2} is defined by
\begin{align}
x_i(k+1)=\max_{j\in \mathcal{N}_i(k)} x_j(k),
\end{align}
so distributed maximizing  is  each node interacting with its neighbors and simply taking  the maximal state within its neighbor set.

In this paper, we aim to present a model  under which we can discuss fundamental  differences of some distributed information processing mechanisms. We consider the following algorithm  for the node updates:
\begin{align}\label{9}
x_i(k+1)=\eta_k x_i(k)+\alpha_k \min_{j\in \mathcal{N}_i(k)}x_j(k)+\big(1-\eta_k-\alpha_k\big)\max_{j\in \mathcal{N}_i(k)} x_j(k),
\end{align}
where $\alpha_k, \eta_k\geq 0$ and $\alpha_k+\eta_k \leq 1$. We denote the set of all algorithms of the form (\ref{9}) by $\mathcal{A}$, when the parameter $(\alpha_k, \eta_k)$ takes value as $ \eta_k\in[0,1], \alpha_k\in[0,1-\eta_k].$ This model is a special case of (\ref{r2}) as the parameter $\alpha_k$ is not depending on the node index $i$ in (\ref{9}).

Note that $\mathcal{A}$ represents a uniform model for distributed averaging and maximizing algorithms. Obeying the cohesion and self-confidence assumptions, the set of (weighted) averaging algorithms, $\mathcal{A}_{\rm ave}$, consists of algorithms in the form of (\ref{9}) with parameters $\eta_k\in(0,1], \alpha_k\in[0,1-\eta_k].$
The set of maximizing algorithms,  $\mathcal{A}_{\rm max}$, is given by the parameter set $  \eta_k\equiv 0\ {\rm and}\ \alpha_k \equiv 0 .$

\begin{remark}
Algorithm (\ref{9}) is more restrictive than (\ref{r2}) in the sense that the averaging weight $\alpha^{\langle i\rangle}_k$ in (\ref{r2}) might vary for different nodes. Hence, (\ref{9}) cannot in general capture the averaging algorithm (\ref{r1}).  Except for this  difference, the standing assumptions  A1 and  A2 of average consensus algorithms are still fulfilled for algorithm (\ref{9}). Moreover, note that for (\ref{9})  no lower bound is imposed on the node self-confidence.
\end{remark}

\begin{remark}
In Algorithm (\ref{9}) each node's update only depends on  the states of the minimum and maximum neighbor states at every time step. In other words, not all links  are  active explicitly in the iterations.  Therefore, the existing convergence results on averaging algorithms cannot be applied  directly,  since these results rely on the connectivity of the communication graph.
\end{remark}

\begin{remark}
If  besides A1 we impose a  double stochasticity assumption \cite{tsi11,tsi2,vb2} on the arc weights $a_{ij}(k)$, i.e., $\sum_{i\in\mathcal{N}_j(k)} a_{ji}(k)=1$ for all $i$ and $k$,  each node's state will converge to the  average of all initial values during the evolution of Algorithm (\ref{r1}) as long as consensus is reached. In the absence of  double stochasticity assumption, if there is a consensus under time-varying communication, we  know that the consensus limit of (\ref{r1})  still is   a convex combination of the initial values with coefficients  invariant  with respect to  initial conditions \cite{caoming1,caoming2,tsi11}. However, neither of these conclusions  holds for  Algorithm (\ref{9}).   It is straightforward to see that the convergence limit is a convex combination of the initial values if consensus is reached. But  due to the state-dependent node update in (\ref{9}),  the coefficients in the convex combination of the consensus limit indeed depend on the initial condition (even with fixed communication graph).
\end{remark}
\subsection{Problem}
 Let $\big\{x(k;x^0)=\big(x_1(k;x^0) \dots x_n(k;x^0)\big)^T\big\}_0^\infty$ be the sequence generated by (\ref{9}) for initial time $k_0$ and initial value  $x^0=x(k_0)=\big(x_1(k_0)\dots x_n(k_0)\big)^T \in\mathds{R}^n$. We will identify $x(k;x^0)$ as $x(k)$ in the following discussions. We introduce the following definition on the convergence of the considered algorithm.
\begin{defn}
(i) Asymptotic consensus is achieved for Algorithm (\ref{9}) for initial condition $x(k_0)=x^0\in\mathds{R}^n$ if there exists $z_\ast(x^0)\in\mathds{R}$ such that
$$
\lim_{k\rightarrow \infty}x_i(k)=z_\ast,\ \ i=1,\dots,n.
$$
Global asymptotic consensus is achieved if asymptotic consensus is achieved for all $k_0\geq 0$ and $x^0\in \mathds{R}^n$.

(ii) Finite-time consensus is achieved for Algorithm (\ref{9}) for  initial condition $x(k_0)=x^0\in\mathds{R}^n$ if there exist $z_\ast(x^0)\in\mathds{R}$ and an integer $T_\ast(x^0)>0$ such that
$$
x_i(T_\ast)=z_\ast,\  \ i=1,\dots,n.
$$
Global finite-time consensus is achieved if finite-time  consensus is achieved for all $k_0\geq 0$ and $x^0\in \mathds{R}^n$.
\end{defn}

The algorithm reaching consensus is equivalent with that $x(k)$ converges to the  manifold
$$
\mathrm{C}=\Big\{x=(x_1 \dots x_n)^T:\ x_1=\dots=x_n\Big\}.
$$
We call  $\mathrm{C}$ the consensus manifold. Its dimension is one.

In the following, we focus on the impossibilities and possibilities of asymptotic or finite-time consensus. We will show that the convergence properties drastically change when Algorithm  (\ref{9}) transits from averaging to maximizing.
\section{Convergence Impossibilities}
In this section, we discuss the impossibilities of asymptotic or finite-time convergence for the averaging algorithms in $\mathcal{A}_{\rm ave}$. One theorem for each case is proven.
\begin{thm}\label{thmr1}
For every averaging algorithm in $\mathcal{A}_{\rm ave}$, finite-time consensus fails for all initial values in $\mathds{R}^n$ except for initial values on the consensus manifold.
\end{thm}
{\it Proof. } We define
$$
h(k)=\min_{i\in\mathcal{V}} x_i(k);\ \ H(k)=\max_{i\in\mathcal{V}} x_i(k).
$$
Introduce $\Phi(k)=H(k)-h(k)$. Then clearly asymptotic consensus is achieved if and only if $\lim_{k\rightarrow \infty} \Phi(k)=0$.

Take a node $i$ satisfying $x_i(k)=h(k)$. We have
\begin{align}\label{11}
x_i(k+1)&=\eta_{k} x_i(k)+\alpha_{k} \min_{j\in \mathcal{N}_i(k)}x_j(k)+\big(1-\eta_{k}-\alpha_{k}\big)\max_{j\in \mathcal{N}_i(k)} x_j(k)\nonumber\\
&\leq   (\alpha_{k}+\eta_{k})h(k)+(1-\eta_{k}-\alpha_{k})H(k).
\end{align}
Similarly, taking another  node $m$ satisfying $x_m(k)=H(k)$, we obtain
\begin{align}\label{12}
x_m(k+1)&=\eta_{k} x_j(k)+\alpha_{k} \min_{m\in \mathcal{N}_i(k)}x_j(k)+\big(1-\eta_{k}-\alpha_{k}\big)\max_{m\in \mathcal{N}_i(k)} x_j(k)\nonumber\\
&\geq   \alpha_{k}h(k)+(1-\alpha_{k})H(k).
\end{align}

With (\ref{11}) and (\ref{12}), we conclude that
\begin{align}\label{14}
\Phi (k+1)&=  \max_{i\in\mathcal{V}} x_i(k)- \min_{i\in\mathcal{V}} x_i(k)\geq   x_m(k+1)-x_i(k+1)\geq \eta_k \Phi(k).
\end{align}
Therefore, since (\ref{14}) holds for all $k$, we immediately obtain that for  every algorithm in the averaging set  $\mathcal{A}_{\rm ave}$,
\begin{align}\label{15}
\Phi(K) \geq \Phi(k_0)\prod_{k=k_0}^{K-1} \eta_k >0
\end{align}
for all $K\geq k_0$ as long as  $\Phi(k_0)>0$. Noticing that the initial values satisfying $\Phi(k_0)=0$ are on the  consensus manifold, the desired conclusion follows. \hfill$\square$

Since the consensus manifold is a one-dimensional manifold in $\mathds{R}^n$,  Theorem \ref{thmr1} indicates that finite-time convergence  is almost impossible for average consensus algorithms. This partially explains why finite-time convergence results are rare for averaging algorithms in the literature.

Next, we discuss  the impossibility of asymptotic consensus.  The following lemma is well-known.
\begin{lem}\label{lemr1}
Let $\{b_k\}_0^\infty$ be a sequence of real numbers with $b_k\in[0,1)$ for all $k$. Then $\sum_{k=0}^\infty b_k=\infty$ if and only if $\prod_{k=0}^{\infty}(1-b_k)=0$.
\end{lem}

The following theorem on asymptotic convergence holds.
\begin{thm}\label{thmr2}
For every averaging algorithm in $\mathcal{A}_{\rm ave}$, asymptotic consensus fails for all initial values in $\mathds{R}^n$ except for initial values on the consensus manifold, if $\sum_{k=0}^\infty \big(1-\eta_k\big)<\infty$.
\end{thm}
{\it Proof.} In light of Lemma \ref{lemr1}  and (\ref{15}), we see that for  every algorithm in the averaging set  $\mathcal{A}_{\rm ave}$,
\begin{align}
\lim_{K\rightarrow \infty}\Phi(K) \geq \Phi(k_0)\prod_{k=k_0}^{\infty} \eta_k >0
\end{align}
if $\sum_{k=0}^\infty \big(1-\eta_k\big)<\infty$ for all initial values satisfying  $\Phi(k_0)>0$. The desired conclusion thus follows. \hfill$\square$

Theorem \ref{thmr2} indicates that $\sum_{k=0}^\infty \big(1-\eta_k\big)=\infty$ is a necessary condition for average algorithms to reach asymptotic consensus.  Note that $\eta_k$ measures node  self-confidence. Thus, the condition $\sum_{k=0}^\infty \big(1-\eta_k\big)=\infty$ characterizes the maximal self-confidence that nodes can hold and still reach consensus.

It is worth pointing out that Theorems  \ref{thmr1} and \ref{thmr2} hold for any communication graph. In the following discussions, we turn to the possibilities for consensus. Then, however,  the communication graph plays an important role. Time-dependent and state-dependent graphs will be studied, respectively, in the following two sections.
\section{Time-dependent Graphs}
In this section, we focus on  time-dependent graphs. We first discuss a special case when the network topology is fixed, and the  required connectivity for average and max-min algorithms will be treated. Next, time-varying communications will be discussed, and a series of conditions will be presented on the asymptotic or finite-time convergence of the considered algorithm under jointly connected graphs.

\subsection{Fixed Graph}
For  fixed communication graphs, we present the following result.
\begin{thm}\label{thmt1}
Suppose $\mathcal{G}_k\equiv\mathcal{G}_\ast$ is a fixed digraph for all $k$.

(i) For every algorithm in  $\mathcal{A}_{\rm ave}$,  global asymptotic consensus can be achieved   only if $\mathcal{G}_\ast$ is quasi-strongly connected.

(ii) For every algorithm in  $\mathcal{A}_{\rm max}$,  global finite-time  consensus is achieved   if and only if $\mathcal{G}_\ast$ is strongly connected.
\end{thm}
{\it Proof.}  (i) If $\mathcal{G}_\ast$ is not quasi-strongly connected, there exist two distinct nodes $i$ and $j$ such that $\mathcal {V}_1\cap \mathcal {V}_2=\emptyset$, where $\mathcal {V}_1=\{\mbox{nodes\ from\ which\ $i$\ is\ reachable\ in}\  \mathcal
{G}_\ast\}$ and  $\mathcal {V}_2=\{$nodes\ from\ which\ $j$\ is\ reachable\ in $\mathcal
{G}_\ast\}$.  Consequently, nodes in $\mathcal{V}_1$ never receive information from nodes in $\mathcal{V}_2$. Take $x_i(k_0)=0$ for $i\in \mathcal{V}_1$ and $x_i(k_0)=1$ for $i\in \mathcal{V}_2$. Obviously, consensus cannot be achieved under this initial condition. The conclusion holds.

(ii) (Sufficiency.) Let $v_0$ be a  node with the maximal value initially. Then after one step  all the nodes for which  $v_0$ is a neighbor will reach the maximal value. Proceeding the analysis we see that the whole network will converge to the initial maximum in finite time.

\noindent (Necessity.) Assume that $\mathcal{G}_\ast$ is not strongly connected. There will be two different nodes $i_{\ast}$ and $j_{\ast}$ such that $j_\ast$ is not reachable from $i_\ast$. Introduce $\mathcal{V}_\ast=\{j:$ $j$ is reachable from $i_\ast\}$. Then $\mathcal{V}_\ast\neq \mathcal{V}$ because $j_\ast\notin \mathcal{V}_\ast$. Moreover, the definition of $\mathcal{V}_\ast$ guarantees that all the nodes in $\mathcal{V}\setminus \mathcal{V}_\ast$ will never be influenced by the nodes in $\mathcal{V}_\ast$. Therefore, letting the initial maximum be unique and reached by some node in $\mathcal{V}_\ast$, consensus will not be reached.

The proof is complete.\hfill$\square$

As will be shown  in the following discussions, quasi-strong connectivity is not only necessary, but also sufficient to guarantee global asymptotic consensus  for the algorithms in the averaging set $\mathcal{A}_{\rm ave}$ under some mild conditions on the parameters $(\alpha_k,\eta_k)$. Therefore, Theorem \ref{thmt1} clearly states  that  quasi-strong connectivity is critical for averaging, as is strong connectivity for  maximizing.

\subsection{Time-varying Graph}
We now turn to time-varying graphs.  Denote the joint graph of $\mathcal
{G}_k$ over
time interval $[k_1,k_2]$  as
$\mathcal {G}\big([k_1,k_2]\big)=(\mathcal {V},\cup_{k\in[k_1,k_2]}\mathcal
{E}(k))$, where $0\leq k_1\leq k_2\leq +\infty$.  We introduce the following definitions on the joint connectivity of time-varying graphs.
\begin{defn}
(i) $\mathcal
{G}_k$ is  {\it uniformly jointly quasi-strongly connected} (strongly connected) if there exists an integer  $B\geq1$ such that  $
\mathcal {G}\big([k,k+B-1]\big)$ is quasi-strongly connected (strongly connected) for all $k\geq0$.

(ii) $\mathcal
{G}_k$ is {\it infinitely jointly strongly connected}  if  $
\mathcal {G}\big([k,\infty)\big)$ is  strongly connected for all $k\geq0$.

(iii) Suppose $\mathcal {G}_k$ is bidirectional for all $k$. Then  $\mathcal
{G}_k$ is  {\it infinitely jointly connected}  if  $
\mathcal {G}\big([k,\infty)\big)$ is  connected for all $k\geq0$.
\end{defn}

\begin{remark}
The uniformly joint connectivity, which requires the union graph to be connected over each bounded interval, has been extensively studied in the literature, e.g., \cite{tsi,jad03,caoming1,caoming2,ren}. The infinitely joint connectivity is a  more general case since it does not impose an upper bound for the length of the interval where connectivity is guaranteed for the union graph. Convergence results for consensus algorithms based on infinitely joint connectivity are given in \cite{mor,shi09,shi11}.
\end{remark}

The following conclusion holds for uniformly jointly quasi-strongly connected graphs.
\begin{thm}\label{thmt2}  Suppose  $\mathcal{G}_k$ is uniformly jointly quasi-strongly connected. Algorithms in the averaging set $\mathcal{A}_{\rm ave}$ achieve global asymptotic consensus
if either
\begin{align}\label{20}
\Huge{\sum}_{s=0}^\infty \bigg[ \prod_{k= s(n-1)^2B}^{(s+1)(n-1)^2B-1}\alpha_k \bigg]=\infty
\end{align}
or
\begin{align}\label{21}
\Huge{\sum}_{s=0}^\infty \bigg[\prod_{k= s(n-1)^2B}^{(s+1)(n-1)^2B-1}\big(1-\alpha_k -\eta_k\big)\bigg] =\infty.
\end{align}
\end{thm}

Theorem \ref{thmt2} hence states that divergence of certain products of the algorithm parameters guarantees global asymptotic consensus.

It is  straightforward to see that  for a non-negative sequence $\{b_k\}$ with $b_k\geq b_{k+1}$ for all $k$, $\sum_{s=0}^\infty \prod_{k= s(n-1)^2B}^{(s+1)(n-1)^2-1}b_k =\infty$ if and only if  $\sum_{k=0}^\infty b_k^{(n-1)^2B} =\infty$. Thus, the following corollary follows from Theorem \ref{thmt2}.

\begin{coro}
Suppose  $\mathcal{G}_k$ is uniformly jointly quasi-strongly connected.

 (i) Assume that  $\alpha_k\geq \alpha_{k+1}$ for all $k$. Algorithms in the averaging set  $\mathcal{A}_{\rm ave}$ achieve global asymptotic consensus
if $\sum_{k=0}^\infty \alpha_k^{(n-1)^2B} =\infty$.

(ii)  Assume that  $\alpha_k+ \eta_k \leq \alpha_{k+1}+\eta_{k+1}$ for all $k$. Algorithms in the averaging set $\mathcal{A}_{\rm ave}$ achieve global asymptotic consensus
if $\sum_{k=0}^\infty \big(1-\alpha_k-\eta_k\big)^{(n-1)^2B} =\infty$.
\end{coro}

For uniformly jointly strongly connected graphs, it turns out  that consensus can be achieved under  weaker conditions on $(\alpha_k,\eta_k)$.
\begin{thm}\label{thmr4}  Suppose  $\mathcal{G}_k$ is uniformly jointly strongly connected. Algorithms in the averaging set $\mathcal{A}_{\rm ave}$ achieve global asymptotic consensus
if either
\begin{align}
\Huge{\sum}_{s=0}^\infty\bigg[ \prod_{k= s(n-1)B}^{(s+1)(n-1)B-1}\alpha_k \bigg]=\infty
\end{align}
or
\begin{align}
\Huge{\sum}_{s=0}^\infty \bigg[ \prod_{k= s(n-1)B}^{(s+1)(n-1)B-1}\big(1-\alpha_k-\eta_k\big)\bigg] =\infty.
\end{align}
\end{thm}

Similarly, Theorem \ref{thmr4} leads to the following corollary.
\begin{coro}
Suppose  $\mathcal{G}_k$ is uniformly jointly strongly connected.

 (i) Assume that  $\alpha_k\geq \alpha_{k+1}$ for all $k$. Averaging algorithms in the  set $\mathcal{A}_{\rm ave}$ achieve global asymptotic consensus
if $\sum_{k=0}^\infty \alpha_k^{(n-1)B} =\infty$.

(ii)  Assume that  $\alpha_k+ \eta_k \leq \alpha_{k+1}+\eta_{k+1}$ for all $k$.  Averaging algorithms in the  set $\mathcal{A}_{\rm ave}$ achieve global asymptotic consensus
if $\sum_{k=0}^\infty \big(1-\alpha_k-\eta_k\big)^{(n-1)B} =\infty$.
\end{coro}

For bidirectional graphs, the conditions are much simpler to state. We present the following result.

\begin{thm}\label{thmr3}
Suppose $\mathcal {G}_k$ is bidirectional for all $k$ and $\mathcal{G}_k$ is infinitely jointly  connected.  Averaging algorithms in the  set $\mathcal{A}_{\rm ave}$ achieves  achieve global asymptotic consensus if there exists  a constant $\alpha_\ast \in (0,1)$ such that  either $\alpha_k\geq  \alpha_\ast$ or $1-\alpha_k-\eta_k\geq  \alpha_\ast$ for all $k$.
\end{thm}

The convergence of algorithms in the maximizing set $\mathcal{A}_{\rm max}$ is stated as follows.
\begin{thm}\label{thmt3}
 Maximizing  algorithms in the set $\mathcal{A}_{\rm max}$ achieve global finite-time  consensus if  $\mathcal{G}_k$ is infinitely jointly strongly connected.
\end{thm}

Theorems \ref{thmt2}--\ref{thmt3} together provide a comprehensive understanding of the convergence conditions for the considered model (\ref{9}) under time-varying graphs.    Infinitely jointly strong connectivity is sufficient for global finite-time consensus for algorithms in $\mathcal{A}_{\rm max}$ according to Theorem \ref{thmt3}, while infinitely joint connectivity cannot ensure global asymptotic consensus for algorithms in  $\mathcal{A}_{\rm ave}$ in general. Thus, in this sense algorithms in $\mathcal{A}_{\rm ave}$ and $\mathcal{A}_{\rm max}$ are fundamentally different under infinitely jointly connected graphs.

The rest of this section contains the proofs of Theorems \ref{thmt2}--\ref{thmt3}.

\subsubsection{\bf Proof of Theorem \ref{thmt2}}
We continue to  use the following  notations:
$$
h(k)=\min_{i\in\mathcal{V}}x_i(k), \quad H(k)=\max_{i \in\mathcal{V}}x_i(k),
$$
and $\Phi(k)=H(k)-h(k)$. Following any solution of (\ref{9}), it is obvious to see that $h(k)$ is non-decreasing and $H(k)$ is non-increasing.

Note that if (\ref{20}) guarantees asymptotic consensus for algorithm (\ref{9}), replacing the node states $x_i(k)$ with $-x_i(k)$  leads to that (\ref{21}) guarantees asymptotic consensus of algorithm (\ref{9}) for $-x_i(k), i=1,\dots,n$. Since  consensus for $x_i(k), i=1,\dots,n$ is equivalent with  consensus for $-x_i(k), i=1,\dots,n$, (\ref{20}) and (\ref{21}) are equivalent in terms of consensus convergence. Thus, we just need to show that (\ref{20}) is a sufficient condition for asymptotic consensus.

 Take $k_\ast\geq0$ as any moment in the iterative algorithm. Take $(n-1)^2$ intervals $[k_\ast,k_\ast+B-1]$, $[k_\ast+B,k_\ast+2B-1], \dots$, $[k_\ast+(n^2-2n)B,k_\ast+(n-1)^2B-1]$. Since  $\mathcal{G}_k$ is uniformly jointly quasi-strongly connected, the union graph on each of these intervals has at least one center node. Consequently, there must be a node $v_0$ and $n-1$ integers  $0\leq b_1<b_2<\dots<b_{n-1}\leq n^2-2n$ such that $v_0$ is a center of $\mathcal{G}\big([k_\ast+b_iB,k_\ast+(b_i+1)B-1]\big),i=1,\dots,n-1$. Assume that
\begin{align}
x_{v_0}(k_\ast)\leq \frac{1}{2} h(k_\ast)+ \frac{1}{2} H(k_\ast).
\end{align}

We first bound $x_{v_0}(k)$ for $k\in[k_\ast,k_\ast+(n-1)^2B]$. It is not hard to see that
\begin{align}
x_{v_0}(k_\ast+1)&=\eta_{k_\ast}x_{v_0}(k_\ast)+ \alpha_{k_\ast} \min_{j\in \mathcal{N}_{v_0}(k_\ast)}x_j(k_\ast)+\big(1-\alpha_{k_\ast}-\eta_{k_\ast}\big)\max_{j\in \mathcal{N}_{v_0}(k_\ast)} x_j(k_\ast)\nonumber\\
&\leq \big(\alpha_{k_\ast}+\eta_{k_\ast}\big) \Big(\frac{1}{2} h(k_\ast)+ \frac{1}{2} H(k_\ast)\Big)+\big(1-\alpha_{k_\ast}-\eta_{k_\ast}\big)H(k_\ast)\nonumber\\
&\leq \alpha_{k_\ast} \Big(\frac{1}{2} h(k_\ast)+ \frac{1}{2} H(k_\ast)\Big)+\big(1-\alpha_{k_\ast}\big)H(k_\ast)\nonumber\\
&= \frac{\alpha_{k_\ast}}{2}h(k_\ast)+\big(1-\frac{\alpha_{k_\ast}}{2}\big)H(k_\ast).
\end{align}
Proceeding,  we obtain
\begin{align}\label{3}
x_{v_0}(k_\ast+m)\leq \frac{\prod_{k=k_\ast} ^{k_\ast +m-1}\alpha_{k} }{2}h(k_\ast)+\big(1-\frac{\prod_{k=k_\ast} ^{k_\ast +m-1}\alpha_{k} }{2}\big)H(k_\ast),\ \ m=0,1,\dots.
\end{align}

Since $v_0$ is a center of $\mathcal{G}\big([k_\ast+b_1B,k_\ast+(b_1+1)B-1]\big)$, there exists another node $v_1$ such that $v_0$ is a neighbor of $v_1$ for some $k_1\in [k_\ast+b_1B,k_\ast+(b_1+1)B-1]$. As a result, based on (\ref{3}), we have
\begin{align}
x_{v_1}({k}_1+1)& =\eta_{k_1}x_{v_1}(k_1)+\alpha_{k_1}\min_{j\in \mathcal{N}_{v_1}(k_1)}x_j(k_1)+\big(1-\alpha_{k_1}-\eta_{k_1}\big)\max_{j\in \mathcal{N}_{v_1}(k_1)} x_j(k_1)\nonumber\\
&\leq  \alpha_{k_1} x_{v_0}(k_1)+\big(1-\alpha_{k_1}\big)H(k_1)\nonumber\\
&\leq  \alpha_{k_1} \Big(\frac{\prod_{k=k_\ast} ^{k_1-1}\alpha_{k} }{2}h(k_\ast)+\big(1-\frac{\prod_{k=k_\ast} ^{k_1-1}\alpha_{k}}{2}\big)H(k_\ast)\Big)+\big(1-\alpha_{k_1}\big)H(k_\ast)\nonumber\\
&= \frac{\prod_{k=k_\ast} ^{k_1}\alpha_{k} }{2}h(k_\ast)+\big(1-\frac{\prod_{k=k_\ast} ^{k_1}\alpha_{k}}{2}\big)H(k_\ast).
\end{align}
Proceeding,  we have
\begin{align}
x_{v_0}(k_\ast+m)\leq \frac{\prod_{k=k_\ast} ^{k_\ast +m-1}\alpha_{k} }{2}h(k_\ast)+\big(1-\frac{\prod_{k=k_\ast} ^{k_\ast +m-1}\alpha_{k} }{2}\big)H(k_\ast),\ \  m=(b_1+1)B,\dots.
\end{align}

Continuing the analysis on time intervals $[k_\ast+b_iB,k_\ast+(b_i+1)B-1]$ for $i=2,\dots,n-1$ and nodes $v_2,v_3,\dots,v_{n-1}$, similar upper bounds for each node can be obtained:
\begin{align}
x_{v_i}(k_\ast+m)\leq \frac{\prod_{k=k_\ast} ^{k_\ast +m-1}\alpha_{k} }{2}h(k_\ast)+\big(1-\frac{\prod_{k=k_\ast} ^{k_\ast +m-1}\alpha_{k} }{2}\big)H(k_\ast),\ \ m=(b_i+1)B,\dots.
\end{align}
This immediately leads to
\begin{align}
x_{v_i}(k_\ast+(n-1)^2B)\leq \frac{\prod_{k=k_\ast} ^{k_\ast +(n-1)^2B-1}\alpha_{k} }{2}h(k_\ast)+\big(1-\frac{\prod_{k=k_\ast} ^{k_\ast +(n-1)^2B-1}\alpha_{k} }{2}\big)H(k_\ast),\ i=0,1,\dots,n
\end{align}
which implies
\begin{align}
H\big(k_\ast+(n-1)^2B\big)\leq \frac{\prod_{k=k_\ast} ^{k_\ast +(n-1)^2B-1}\alpha_{k} }{2}h(k_\ast)+\big(1-\frac{\prod_{k=k_\ast} ^{k_\ast +(n-1)^2B-1}\alpha_{k} }{2}\big)H(k_\ast).
\end{align}
Thus, we have
\begin{align}\label{6}
\Phi\big(k_\ast+(n-1)^2B\big)&= {H}\big(k_\ast+(n-1)^2B\big)-{h}\big(k_\ast+(n-1)^2B\big)\nonumber\\
&\leq \frac{\prod_{k=k_\ast} ^{k_\ast +(n-1)^2B-1}\alpha_{k} }{2}h(k_\ast)+\big(1-\frac{\prod_{k=k_\ast} ^{k_\ast +(n-1)^2B-1}\alpha_{k} }{2}\big)H(k_\ast)-h(k_\ast)\nonumber\\
&=\Big(1- \frac{\prod_{k=k_\ast} ^{k_\ast +(n-1)^2B-1}\alpha_{k} }{2}\Big)\Phi(k_\ast).
\end{align}

From a symmetric analysis by bounding $h(k_\ast+(n-1)^2B)$ from below, we know that (\ref{6}) also holds for the other condition with $x_{v_0}(k_\ast)\geq \frac{1}{2} h(k_\ast)+ \frac{1}{2} H(k_\ast)$. Therefore, since $k_\ast$ is selected arbitrarily, we can assume the initial time is $k_0=0$, without loss of generality,  and then conclude that
\begin{align}
\Phi\big(K(n-1)^2B\big)
&\leq \Phi(0) \prod_{s=0} ^{K-1}\Bigg(1- \frac{1 }{2}\prod_{k=s(n-1)^2B} ^{(s+1)(n-1)^2B-1}\alpha_{k}\Bigg).
\end{align}

The desired conclusion follows immediately from Lemma \ref{lemr1}.

\subsubsection{Proof of Theorem \ref{thmr4}}
Notice that in a strongly connected graph, every node is a center node. Therefore,  when $\mathcal{G}_k$ is uniformly jointly strongly connected, taking $k_\ast\geq0$ as any moment in the iteration and  $n-1$ intervals $[k_\ast,k_\ast+B-1]$, $[k_\ast+B,k_\ast+2B-1], \dots$, $[k_\ast+(n-2)B,k_\ast+(n-1)B-1]$, any node $i\in\mathcal{V}$ is a center node for  the union graph over each of these intervals. As a result,  the desired conclusion follows repeating  the  analysis used in the proof of Theorem \ref{thmt2}.
\subsubsection{Proof of Theorem \ref{thmr3}}
Similar to the proof of Theorem \ref{thmt2},  we only need to show that the existence of  a constant $\alpha_\ast \in (0,1)$ such that  $\alpha_k\geq  \alpha_\ast$ is sufficient for asymptotic consensus.

  Take $k^\ast_1\geq0$ as an arbitrary  moment in the iterative algorithm. Take a node $u_0\in\mathcal{V}$ satisfying $x_{u_0}(k^\ast_1)=h(k^\ast_1)$. We define
\begin{align}
k_1= \inf\big\{k\geq k^\ast_1:\mbox{there exists another node connecting $u_0$ at time}\ k\big\}
\end{align}
and then
\begin{align}
\mathcal{V}_1= \big\{k\geq k^\ast_1:\mbox{nodes which are connected to $u_0$ at time}\ k_1\big\}.
\end{align}
Thus, we have
\begin{align}
x_{u_0}(k_1+1)&=\eta_{k_1}x_{u_0}(k_1)+\alpha_{k_1}\min_{j\in \mathcal{N}_{u_0}(k_1)}x_j(k_1)+\big(1-\alpha_{k_1} -\eta_{k_1}\big)\max_{j\in \mathcal{N}_{u_0}(k_1)} x_j(k_1)\nonumber\\
&\leq  (\alpha_{k_1}+\eta_{k_1}) x_{u_0}(k_1)+\big(1-\alpha_{k_1}-\eta_{k_1}\big)H(k_1)\nonumber\\
&\leq  (\alpha_{k_1}+\eta_{k_1}) h(k_1^\ast)+\big(1-\alpha_{k_1}-\eta_{k_1}\big)H(k_1^\ast)\nonumber\\
&\leq  \alpha_{k_1} h(k_1^\ast)+\big(1-\alpha_{k_1}\big)H(k_1^\ast)\nonumber\\
&\leq \alpha_{\ast} h(k_1^\ast)+\big(1-\alpha_{\ast}\big)H(k_1^\ast)
\end{align}
and
\begin{align}
x_{i}(k_1+1)&=\eta_{k_1}x_{i}(k_1)+\alpha_{k_1}\min_{j\in \mathcal{N}_{i}(k_1)}x_j(k_1)+\big(1-\alpha_{k_1}\big)\max_{j\in \mathcal{N}_{i}(k_1)} x_j(k_1)\nonumber\\
&\leq  \alpha_{k_1} x_{u_0}(k_1)+\big(1-\alpha_{k_1}\big)H(k_1)\nonumber\\
&\leq \alpha_{k_1} h(k^\ast_1)+\big(1-\alpha_{k_1}\big)H(k^\ast_1)\nonumber\\
&\leq \alpha_{\ast} h(k_1^\ast)+\big(1-\alpha_{\ast}\big)H(k_1^\ast)
\end{align}
for all $i\in \mathcal{V}_1$.

Note that if nodes in $\{u_0\}\cup\mathcal{V}_1$ are not connected with other nodes in $\mathcal{V}\setminus (\{u_0\}\cup\mathcal{V}_1)$ during $[k_1+1,k_1+s]$, $s\geq 1$, we have that for all $i\in\{u_0\}\cup \mathcal{V}_1$,
 \begin{align}
x_{i}(k_1+m)\leq \alpha_{\ast} h(k_1^\ast)+\big(1-\alpha_{\ast}\big)H(k_1^\ast),\ \ m=1,\dots,s+1.
\end{align}

Continuing the estimate,  $k_2,\dots,k_d$ and $\mathcal{V}_2,\dots,\mathcal{V}_d$ can be defined correspondingly until $\mathcal{V}=\{u_0\}\cup(\cup_{i=1}^d \mathcal{V}_i)$, so eventually we have
 \begin{align}
x_{i}(k_d+1)\leq \alpha^d_\ast h(k^\ast_1)+\big(1-\alpha^d_\ast\big)H(k^\ast_1), \ \ i=1,\dots,n,
\end{align}
which implies
\begin{align}\label{5}
{H}(k_d+1)\leq \alpha^d_\ast h(k^\ast_1)+\big(1-\alpha^d_\ast\big)H(k^\ast_1).
\end{align}
We denote $k^\ast_2=k_d+1$. Because it holds that $d\leq n-1$, we see from (\ref{5}) that
\begin{align}
\Phi(k^\ast_2)\leq \big(1-\alpha^{n-1}_\ast\big)\Phi(k^\ast_1).
\end{align}
Since $\mathcal{G}_k$ is infinitely jointly  connected, this process can be carried on for an infinite sequence $k_1^\ast<k_2^\ast<\dots$. Thus, asymptotic consensus is achieved for all initial conditions.  This completes the proof.

\subsubsection{Proof of Theorem \ref{thmt3}}
 Let $v_0$ be a  node with the maximal value initially. Because $\mathcal{G}_k$ is infinitely jointly strongly connected,  we  can  define
\begin{align}
k_1= \inf\big\{k\geq k^\ast_1:\mbox{there exists another node for which $v_0$ is a neighbor at time}\ k\big\}
\end{align}
and then
\begin{align}
\mathcal{V}_1= \big\{k\geq k^\ast_1:\mbox{nodes for which $v_0$ is a neighbor at time}\ k_1\big\}.
\end{align}
Then at time ${k}_1+1$ all the nodes in $\mathcal{V}_1$ will reach the maximal value. Proceeding the analysis we know that the whole network will converge to the initial maximum in finite time.\hfill$\square$

\section{State-dependent Graphs}
In this section, we investigate the convergence of Algorithm (\ref{9}) for state-dependent graphs. We are interested in a particular set of averaging algorithms, $\mathcal{A}_{\rm ave}^\ast$, where $(\alpha_k, \eta_k)$ takes value $ \eta_k \equiv 0$, $\alpha_k\in(0,1) $. Algorithms in $\mathcal{A}_{\rm ave}^\ast$ correspond to the case when the self-confidence assumption A2 does not hold, and are  of the form
\begin{align}\label{r100}
x_i(k+1)=\alpha_k\min_{j\in \mathcal{N}_i(k)}x_j(k)+\big(1-\alpha_k\big)\max_{j\in \mathcal{N}_i(k)} x_j(k).
\end{align}
Algorithms in $\mathcal{A}_{\rm ave}^\ast$  still have local cohesion. Hence, they fulfill Assumption A1 but not A2. In fact,   averaging  algorithms without self-confidence have been investigated in classical works on the convergence of product of stochastic matrices, e.g., \cite{wolf,haj,degroot}.
\subsection{State-dependent  Communication}

 In both Krause's \cite{krause}  and Vicsek's \cite{vic95} models, nodes interact with neighbors whose distance is within a certain communication range. Convergence analysis for consensus algorithms under such models  can be found in \cite{vb1,julien,tang,liuguo}. Recently, it was discovered  through empirical data that in a bird flock  each bird seems to  interact with a fixed number of nearest neighbors, rather than with all neighbors within a fixed metric distance \cite{pnas}. Nearest-neighbor model has been  studied under a probabilistic setting on the graph  connectivity for wireless communication networks \cite{kumar}. From a social network point of view, the evolution of  opinions may  result from similar models since members  tend to exchange information  with a fixed number of other members  who hold a similar opinion as themselves \cite{degroot,julien}. In this section, we consider a network model in which nodes interact only with other  nodes having a close state value. Consider the following  nearest-neighbor rule.

\begin{defn}[Nearest-neighbor Graph]   For  a positive integer $\mu$ and any node $i\in\mathcal{V}$, there is a link entering $i$ from each node in the set $\mathcal{N}_i^-(k)\cup\mathcal{N}_i^+(k)$, where $$
\mbox{$\mathcal{N}_i^-(k)=\big\{$nearest $\mu$ neighbors from $\{j\in\mathcal{V}: x_{j}(k)$$<x_i(k)\}\big\}$}
$$
denotes the nearest smaller neighbor set, and $$
\mbox{$\mathcal{N}_i^+(k)=\big\{$nearest $\mu$ neighbors from $\{j\in\mathcal{V}: x_{j}(k)$$>x_i(k)\}\big\}$}
 $$
 denotes the nearest larger neighbor set. The graph defined by this nearest neighbor rule is denoted as $\mathcal{G}^{\mu}_{x(k)}$, $k=0,1,\dots$.
\end{defn}

 Naturally, if there are less than $\mu$ nodes with states smaller than $x_i(k)$, $\mathcal{N}_i^-(k)$ has less that $\mu$ elements. Similar condition holds for  $\mathcal{N}_i^+(k)$. Hence, the number of neighbor nodes is not necessarily fixed in the nearest-neighbor graph.

\begin{remark}
Note that, at each time $k$, the nearest-neighbor graph is uniquely determined by the  node states.  The node interactions are indeed determined  by the distance between the node states. In this sense, the nearest-neighbor graph shares similar structure with  Krause's model \cite{krause,vb1}, where each node communicates with the nodes within certain radius. This nearest-neighbor graph also fulfills the interaction structure in the bird flock model discussed in \cite{pnas} since each node communicates with an almost fixed number of neighbors,  nearest from above and below.
\end{remark}

Note that in the definition of the nearest-neighbor graph, nodes may have neighbors with the same state values. We consider the following nearest-value graph, where each node considers only neighbors with different state values.

  \begin{defn}{\it (Nearest-value Graph)}    For  a positive integer $\mu$ and any node $i\in\mathcal{V}$, there is a link entering $i$ from each node in the set $\mathcal{N}_i^-(k)\cup\mathcal{N}_i^+(k)$, where $$
  \mbox{$\mathcal{N}_i^-(k)=\big\{$nearest $\mu$ neighbors with different values from $\{j\in\mathcal{V}: x_{j}(k)$$<x_i(k)\}\big\}$}
  $$
  denotes the nearest smaller neighbor set, and $$
  \mbox{$\mathcal{N}_i^+(k)=\big\{$nearest $\mu$ neighbors with different values from $\{j\in\mathcal{V}: x_{j}(k)$$>x_i(k)\}\big\}$}
   $$
   denotes the nearest larger neighbor set. The graph defined by this nearest neighbor rule is denoted as $\mathcal{G}^{\mu v}_{x(k)}$, $k=0,1,\dots$.
\end{defn}

\begin{figure}
\begin{minipage}[t]{0.45\linewidth}
\centering
\includegraphics[width=\textwidth]{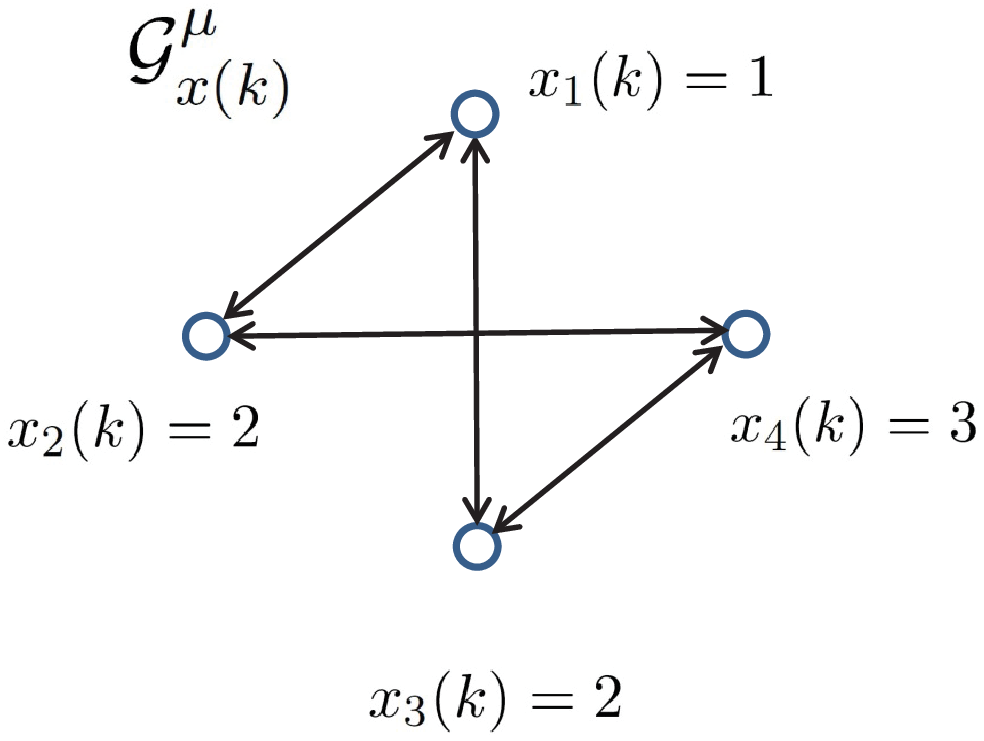}
\end{minipage}
\hfill
\begin{minipage}[t]{0.45\linewidth}
\centering
\includegraphics[width=\textwidth]{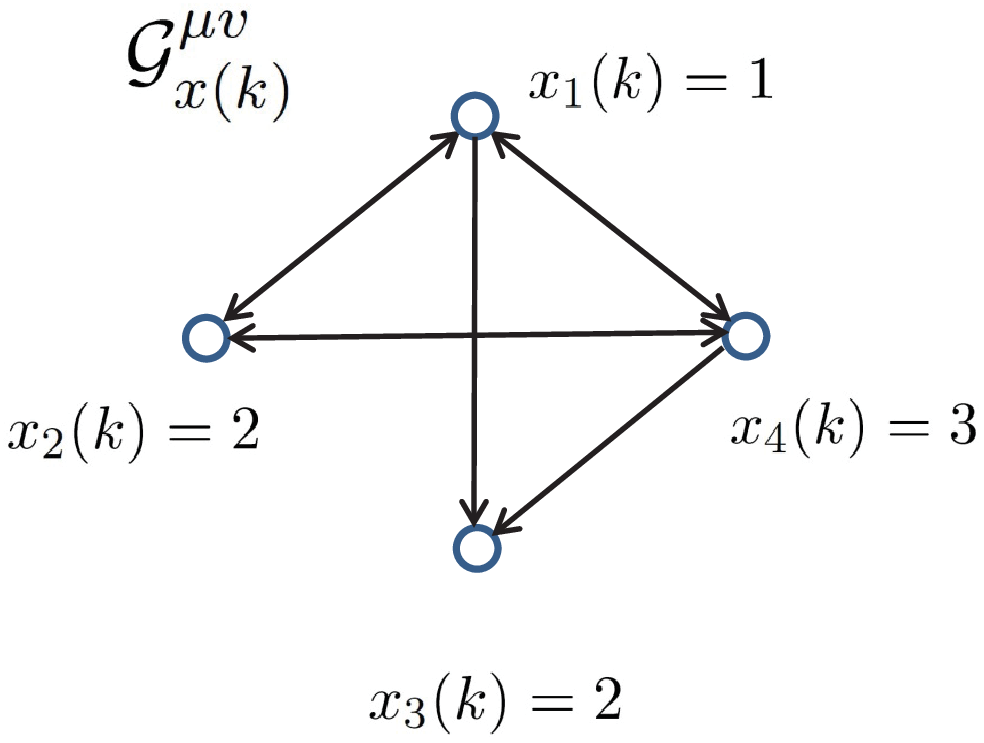}
\end{minipage}
\caption{Examples of nearest-neighbor graph $\mathcal{G}^{\mu}_{x(k)}$ and nearest-value graph $\mathcal{G}^{\mu v}_{x(k)}$ for $\mu=2$. Note that for a given set of states, these graphs are in general not unique.  }\label{diff}
\end{figure}

An illustration of  nearest-neighbor  and nearest-value graphs  at a specific time instance $k$ is shown in Figure \ref{diff} for $n=4$ nodes and $\mu=2$.

\subsection{Basic Lemmas}
We first establish two  useful lemmas for the analysis of nearest-neighbor and nearest-value graphs. The following lemma indicates that the order of node states is preserved.

\begin{lem}\label{lem1}
For any two nodes $u,v\in\mathcal{V}$ and every algorithm in $\mathcal{A}$, under either the nearest-neighbor graph $\mathcal{G}^{\mu}_{x(k)}$ or the nearest-value graph $\mathcal{G}^{\mu v}_{x(k)}$, we have

(i) $x_u(k+1)=x_v(k+1)$ if $x_u(k)=x_v(k)$;

(ii) $x_u(k+1)\leq x_v(k+1)$ if $x_u(k)< x_v(k)$.
\end{lem}
{\it Proof.} When $x_u(k)=x_v(k)$, we have $\{j: x_{j}(k)<x_u(k)\}$=$\{j: x_{j}(k)<x_v(k)\}$ and $\{j: x_{j}(k)>x_u(k)\}$=$\{j: x_{j}(k)>x_v(k)\}$. Thus, for either $\mathcal{G}^{\mu}_{x(k)}$ or $\mathcal{G}^{\mu v}_{x(k)}$, both
$$
\min_{j\in \mathcal{N}_u(k)}  x_j(k) =\min_{j\in \mathcal{N}_v(k)}  x_j(k) \ \  \mbox{and}\ \ \max_{j\in \mathcal{N}_u(k)}  x_j(k) =\max_{j\in \mathcal{N}_v(k)}  x_j(k)
$$ hold. Then (i) follows straightforwardly.

 If $x_u(k)<x_v(k)$, it is easy to see that
$$
\min_{j\in \mathcal{N}_u(k)}  x_j(k)  \leq \min_{j\in \mathcal{N}_v(k)}  x_j(k) ;\quad \max_{j\in \mathcal{N}_u(k)}  x_j(k)  \leq \max_{j\in \mathcal{N}_v(k)}  x_j(k)
$$
according to the definition of neighbor sets, which implies (ii) immediately. \hfill$\square$

Define
$$
\Upsilon_k=\Big| \big\{x_1(k),\dots,x_n(k)\big\}\Big|
$$
as the number of distinct node states at time $k$, where $\big|S\big|$ for a set $S$ represents its cardinality. Then Lemma \ref{lem1} implies that $\Upsilon_{k+1}\leq \Upsilon_k$ for all $k\geq 0$. This point plays an important role  in the convergence analysis.

Moreover,   for both the nearest-neighbor graph $\mathcal{G}^\mu_{x(k)}$ and the  nearest-value graph $\mathcal{G}^{\mu v}_{x(k)}$, in order to distinguish the node states under different values of neighbors, we denote $x_i^\mu(k)$ as the state of node $i$ when the number of larger or smaller neighbors is $\mu$.  Correspondingly, we denote $$
h^{\mu}(k)=\min_{i\in\mathcal{V}} x_i^\mu (k),\ \ H^\mu(k)=\max_{i\in\mathcal{V}} x_i^\mu(k).
$$
and $\Phi^\mu(k)=H^\mu(k)-h^\mu(k)$. We give another lemma indicating that the convergence speed increases as the number of neighbors increases, which is quite intuitive because  apparently graph connectivity increases as the number of neighbors increases.
\begin{lem}\label{lem2}
Consider  either the nearest-neighbor graph $\mathcal{G}^\mu_{x(k)}$ or the  nearest-value graph $\mathcal{G}^{\mu v}_{x(k)}$. Given two integers $1\leq\mu_1\leq \mu_2$.  For  every algorithm in $\mathcal{A}$ and every initial value,  we have $\Phi^{\mu_1}(k)\geq  \Phi^{\mu_2}(k)$ for all $k$.
\end{lem}
{\it Proof.} Fix the initial condition at time $k_0$. Let $m\in\mathcal{V}$ be a node satisfying $x_m^{\mu_1}(k_0)=h^{\mu_1}(k_0)$ and $x_m^{\mu_2}(k_0)=h^{\mu_2}(k_0)$. The order preservation property given by Lemma \ref{lem1} guarantees that  $x_m^{\mu_1}(k)=h^{\mu_1}(k)$ and $x_m^{\mu_2}(k)=h^{\mu_2}(k)$ for all $k\geq k_0$. It is straightforward to see that  $x_m^{\mu_1}(k_0+1)\leq x_m^{\mu_2}(k_0+1)$ if $\mu_1 \leq \mu_2$, and continuing we know that $x_m^{\mu_1}(k_0+s)\leq x_m^{\mu_2}(k_0+s)$ for all $s\geq 2$.   Thus, we have $h^{\mu_1}(k) \leq h^{\mu_2}(k)$ for all $k\geq k_0$. A symmetric analysis leads to  $H^{\mu_1}(k) \geq H^{\mu_2}(k)$  for all $k$ and the desired conclusion thus follows. \hfill$\square$

\subsection{Convergence for Nearest-neighbor Graph}
For algorithms in the set $\mathcal{A}_{\rm ave}^\ast$, we present the following result under nearest-neighbor graph.
\begin{thm}\label{thm8} Consider the nearest-neighbor graph $\mathcal{G}^\mu_{x(k)}$.

(i) When $n\leq \mu+1$,  each algorithm in  $\mathcal{A}_{\rm ave}^\ast$ achieves global finite-time consensus;

(ii) When $n> \mu+1$, each algorithm in $\mathcal{A}_{\rm ave}^\ast$  fails to achieve finite-time   consensus for almost all initial values;

(iii)  When $n> \mu+1$, each algorithm in $\mathcal{A}_{\rm ave}^\ast$   achieves global asymptotic  consensus  if  $\{\alpha_k\}$ is monotone.
\end{thm}
{\it Proof.} (i) When $n\leq \mu+1$, the communication graph is the complete graph. Thus, consensus will be achieved in one step following (\ref{9}) for every algorithm in $\mathcal{A}_{\rm ave}^\ast$.

(ii) Let $n> \mu+1$. We define two index set
 $$
 \mathcal{I}_k^-=\big\{i:\ x_i(k)=h(k)=\min_{i\in\mathcal{V}} x_i (k)\big\};\   \mathcal{I}_k^+=\big\{i:\ x_i(k)=H(k)=\max_{i\in\mathcal{V}} x_i (k)\big\}.
 $$

 {\it Claim.} Suppose both $\mathcal{I}_k^-$ and  $\mathcal{I}_k^+$ contain one node only. Then so do $\mathcal{I}_{k+1}^-$ and  $\mathcal{I}_{k+1}^+$.

Let  $u$ and $v$ be the unique element in  $\mathcal{I}_k^-$ and  $\mathcal{I}_k^+$, respectively. Take $m\in\mathcal{V}\setminus\{u\}$. Noting the fact that $x_m(k)>x_u (k)$ and $\mu\leq n-2$, we have
$$
\min_{j\in \mathcal{N}_u(k) }  x_j(k) \leq \min_{j\in \mathcal{N}_m(k) }  x_j(k),\ \max_{j\in \mathcal{N}_u(k) }  x_j(k) <\max_{j\in \mathcal{N}_m(k) }  x_j(k).
$$
This leads to $x_{m}(k+1)>x_u(k+1)$. Therefore, $u$ is still the unique element in  $\mathcal{I}_{k+1}^-$. Similarly we can prove that $v$ is still the unique element in  $\mathcal{I}_{k+1}^+$. The claim holds.

Now observe that
$$
\Delta\doteq\bigcup_{u\neq v} \big\{x=(x_1\dots x_n)^T:\ x_u <\min_{m\in\mathcal{V}\setminus\{u\}} x_m\ {\rm and}\ x_v>\max_{m\in\mathcal{V}\setminus\{v\}} x_m,\big\}
$$
has measure zero with respect to the standard Lebesgue measure on $\mathds{R}^n$. For any initial value not  in  $\Delta$, we have both $\mathcal{I}_k^-$ and  $\mathcal{I}_k^+$ contain one unique element, and thus finite-time consensus is impossible. The desired conclusion  follows.

(iii) Recall that
$$
\Upsilon_k=\Big| \big\{x_1(k),\dots,x_n(k)\big\}\Big|.
$$
Since $\Upsilon_{k+1}\leq \Upsilon_k$ holds for all $k$ according to Lemma \ref{lem1}, there exists two integers $0\leq m\leq n$ and $T\geq0$ such that
\begin{align}
\Upsilon_k=m,
\end{align}
for all $k\geq T$.  Thus, we can sort the possible  node states for all $k\geq T$  as
$$
y_1(k)<y_2(k)<\dots< y_m(k).
$$

 Apparently $m\neq 1, 2$ since otherwise the graph is complete  for time $\ell$ with $\Upsilon_{\ell}=1,2$ and consensus is reached after one step. We assume $m\geq 3$ in the following discussions.

Algorithm (\ref{r100}) can be  equivalently transformed to the dynamics on $\{y_1(k),\dots,y_m(k)\}$. Moreover, based on Lemma \ref{lem2}, we only need to prove asymptotic consensus for the case $\mu=1$.

Let $\mu=1$ and $k\geq T$. For  algorithms in $\mathcal{A}_{\rm ave}^\ast$, the dynamics of $\{y_1(k),\dots,y_m(k)\}$ can be written:
\begin{equation}\label{22}
\begin{cases}
y_1(k+1)=\alpha_k y_1(k)+(1-\alpha_k ) y_{2}(k);  \\
y_2(k+1)=\alpha_k y_{1}(k)+(1-\alpha_k) y_{3}(k);\\
\quad \quad \quad  \vdots \\
y_{m-1}(k+1)=\alpha_k y_{m-2}(k)+(1-\alpha_k) y_{m}(k);\\
y_m(k+1)=\alpha_k y_{m-1}(k)+(1-\alpha_k) y_{m}(k).
\end{cases}
\end{equation}

Now let $\{\alpha_k\}$ be monotone, say, non-decreasing. Then we have $\alpha_k \geq \alpha_T>0$. Therefore, for all $k\geq T$, we have
\begin{align}
y_1(k+1)=\alpha_k y_1(k)+(1-\alpha_k ) y_{2}(k)\leq  \alpha_T y_1(k)+(1-\alpha_T) y_{m}(k),
\end{align}
and continuing we know that
\begin{align}
y_1(k+s)\leq  \alpha_T^s y_1(k)+(1-\alpha_T^s) y_{m}(k), s\geq 1.
\end{align}
Similarly for $y_2(k)$, we have
\begin{align}
y_2(k+2)&=\alpha_{k+1}y_{1}(k+1)+(1-\alpha_{k+1}) y_{3}(k+1) \nonumber\\
&\leq  \alpha_T \big( \alpha_T y_1(k)+(1-\alpha_T) y_{m}(k) \big)+(1-\alpha_T) y_{m}(k)\nonumber\\
&\leq  \alpha_T^2 y_1(k)+(1-\alpha_T^2) y_{m}(k)
\end{align}
and
\begin{align}
y_2(k+s)\leq  \alpha_T^s y_1(k)+(1-\alpha_T^s) y_{m}(k), \ s\geq 2.
\end{align}

Proceeding the analysis, eventually we arrive at
\begin{align}
y_i(k+n-1)\leq  \alpha_T^{n-1} y_1(k)+(1-\alpha_T^{n-1}) y_{m}(k), \ i=1,\dots,n,
\end{align}
which yields
\begin{align}
\Phi(k+n-1)&=y_m(k+n-1)-y_1(k+n-1)\nonumber\\
&\leq\alpha_T^{n-1} y_1(k)+(1-\alpha_T^{n-1}) y_{m}(k)-y_1(k)\nonumber\\
&=  (1-\alpha_T^{n-1})\Phi(k).
\end{align}
Thus, global asymptotic consensus is achieved. The other case with $\{\alpha_k\}$ being non-increasing can be proved using a symmetric argument. The desired conclusion follows.




This completes the proof of the theorem.\hfill$\square$

\begin{remark}
In Theorem \ref{thm8}, the asymptotic consensus statement relies on the condition  that $\{\alpha_k\}$ is monotone. From the proof of Theorem \ref{thm8} we see that this condition can be replaced by that  there exists  a constant $\varepsilon \in (0,1)$ such that  either $\alpha_k\geq  \varepsilon$ or $\alpha_k\leq1- \varepsilon$ for all $k$. In fact,  we conjecture that  the asymptotic consensus statement  of Theorem~\ref{thm8} holds true for all $\{\alpha_k\}$, i.e., we believe that asymptotic consensus is achieved for all algorithms in $\mathcal{A}_{\rm ave}^\ast$ under nearest-neighbor graphs.
\end{remark}

\begin{remark}
Theorem \ref{thm8} indicates that $\mu+1$ is a critical number of nodes for nearest-neighbor graphs: for algorithms in $\mathcal{A}_{\rm ave}^\ast$,  finite-time consensus holds  globally if $n\leq \mu+1$, and  fails almost globally if $n>\mu+1$. Note that  $n\leq \mu+1$ implies that the communication  graph is the complete graph, which is rare in general. Recalling that Theorem \ref{thmr1} showed that finite-time consensus fails almost globally for algorithms in $\mathcal{A}_{\rm ave}$, we conclude that finite-time consensus is generally  rare  for averaging  algorithms in $\mathcal{A}$,  no matter with ($\mathcal{A}_{\rm ave}$) or without ($\mathcal{A}_{\rm ave}^\ast$) the  self-confidence assumption.
\end{remark}

For algorithms in $\mathcal{A}_{\rm max}$, we present the following result.
\begin{thm}
 Consider the nearest-neighbor graph $\mathcal{G}^\mu_{x(k)}$. Algorithms in $\mathcal{A}_{\rm max}$ achieve global finite-time  consensus in no more than $\lceil \frac{n}{\mu}\rceil$ steps,  where $\lceil z\rceil$ represents the smallest integer no smaller than $z$.
\end{thm}
{\it Proof.}  Without loss of generality, we assume that $x_1(0),\dots,x_n(0)$ are mutually different. We sort the initial values of the nodes as
$$
x_{i_1}(0)<x_{i_2}(0)<\dots< x_{i_n}(0).
$$
Here $i_m$ denotes node with the $m$'th largest value initially.

Observing that $i_n$ is a right-hand side neighbor of nodes $i_{n-\mu},i_{n-\mu+1},\dots,i_{n-1}$, we have
$$
x_{i_\tau}(1)=x_{i_n}(0), \ \tau=n-\mu,\dots,n.
$$
This leads to $\Upsilon_1= \Upsilon_0-\mu$ where $
\Upsilon_k=\big| \big\{x_1(k),\dots,x_n(k)\big\}\big|$. Proceeding the same analysis we know that consensus is achieved in no more than $\lceil \frac{n}{\mu}\rceil$ steps. The desired conclusion follows. \hfill$\square$
\subsection{Convergence for Nearest-value Graph}
 In this subsection, we study the convergence for nearest-value graphs. Since nearest-value graph $\mathcal{G}^{\mu v}_{x(k)}$ indeed increases the connectivity of  $\mathcal{G}^{\mu}_{x(k)}$, the asymptotic consensus statement of Theorem~\ref{thm8} also holds  for   $\mathcal{G}^{\mu v}_{x(k)}$. The main result for  finite-time consensus of nearest-value graphs is presented as follows. It turns out that the critical number of nodes   for  nearest-value graphs   is  $2\mu$.
\begin{thm}\label{thm10}Consider the nearest-value graph $\mathcal{G}^{\mu v}_{x(k)}$.

 (i) When $n\leq  2\mu$, algorithms in  $\mathcal{A}_{\rm ave}^\ast$ achieve global finite-time consensus in no more than $\lceil \log_2(2\mu+1)\rceil$ steps;

  (ii) When $n> 2\mu$,   algorithms in  $\mathcal{A}_{\rm ave}^\ast$ fail to achieve finite-time consensus for almost all initial conditions.
\end{thm}
{\it Proof.} (i) Suppose $n\leq 2\mu$. Based on Lemma \ref{lem2}, without loss of generality, we assume $n=2\mu$ and the initial values of the nodes are mutually different. Now we have $\Upsilon_0=\big| \{x_1(0),\dots,x_n(0)\}\big|=2\mu$. We fist show the following claim.

\vspace{2mm}
\noindent {\it Claim.} If $\Upsilon_k=2\mu-A$ with $A\geq 0$ an integer, then $\Upsilon_{k+1}\leq \Upsilon_k-A-1$.

\noindent We order the node states at time $k$ and denote them  as
$$
Y_1<Y_2<\dots< Y_{\Upsilon_k}.
$$
When $\Upsilon_k=2\mu-A$, it is not hard to find that the for all $m=\mu-A,\dots,\mu+1$, each node with value $Y_{\Upsilon_m}$ will connect to some node with value $Y_1$, and some other node with value $Y_{\Upsilon_k}$. Therefore, the nodes with value $Y_{\Upsilon_m}, m=\mu-A,\dots,\mu+1$ will reach the same state after the $k$'th update. The claim holds.
\vspace{2mm}

Therefore, by induction we have $\Upsilon_{k}=\max\{0, \Upsilon_0-\sum_{m=0}^{k-1} 2^m\}=\max\{0, 2\mu-(2^k-1)\}$. The conclusion (i) follows straightforwardly.

(ii) Suppose $n>2\mu$. Let $x_1(0),\dots,x_n(0)$ be mutually different. Then it is not hard to see that for any two nodes $u$ and $v$ with $x_u(0)<x_v(0)$, at least one of
$$
\min_{j\in \mathcal{N}_u(0) }  x_j(0) <\min_{j\in \mathcal{N}_v(0) }  x_j(0)
$$
or
$$
\max_{j\in \mathcal{N}_u(0) }  x_j(0) <\max_{j\in \mathcal{N}_v(0) }  x_j(0)
$$
holds. This immediately leads to $x_u(1)<x_v(1)$. Because $u$ and $v$ are arbitrarily chosen, we can  conclude that  $\Upsilon_1=\Upsilon_0$. By an induction argument we see that $\Upsilon_k=\Upsilon_0=n$ for all $k\geq0$, or equivalently, consensus cannot be achieved in finite time. Now observe that
$$
\bigcup_{i\geq j} \big\{x=(x_1\dots x_n)^T:x_i=x_j\big\}
$$
has measure zero with respect to the standard Lebesgue measure on $\mathds{R}^n$.  The desired conclusion thus follows. \hfill$\square$



\section{Numerical Examples}
In this section, we present two numerical examples.

\begin{figure}
\centerline{\epsfig{figure=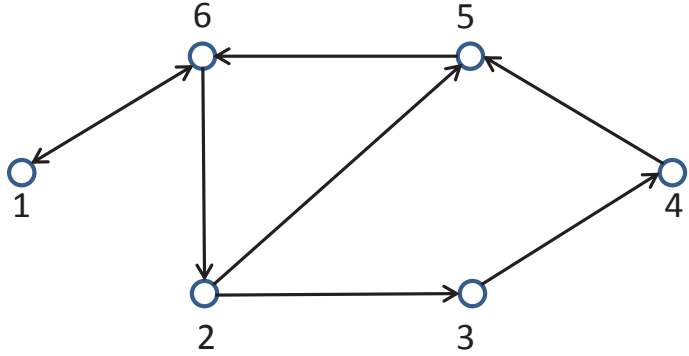, width=0.6\linewidth=0.25}}
\caption{Directed communication graph.}\label{graph}
\end{figure}

\noindent {\bf Example 1.} Consider a network with six nodes $\mathcal{V}=\{1,\dots,6\}$. The communication graph is fixed and directed, as indicated in Figure \ref{graph}.  Initial values for each node are $x_i(0)=i-1, i=1,\dots,6$.

\begin{figure}
\centerline{\epsfig{figure=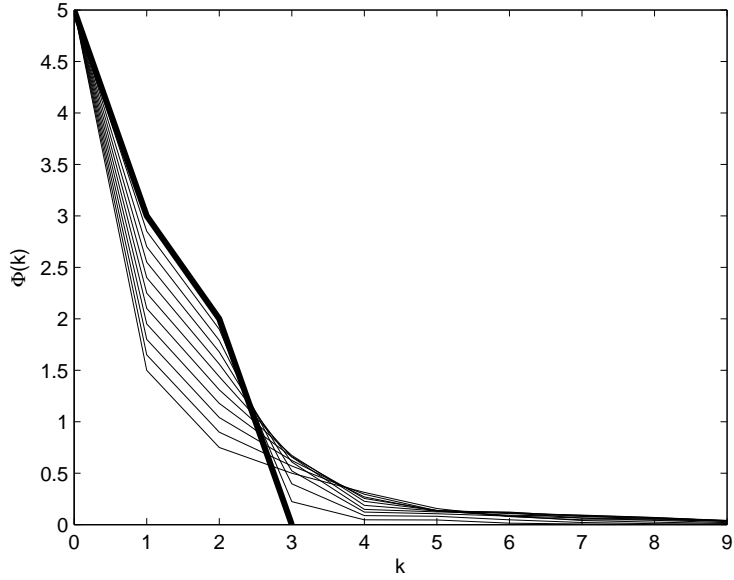, width=0.7\linewidth=0.25}}
\caption{Convergence of Algorithm (\ref{9}) for $\alpha=0,0.05,0.1,\dots,0.5$. The bold line indicates the trajectory with $\alpha=0$, which has finite-time convergence. All other cases have asymptotic convergence. }\label{traj}
\end{figure}

Let $\eta_k\equiv 0$ and $\alpha_k\equiv\alpha$ in Algorithm (\ref{9}). Consider eleven values in the interval  $[0,1/2]$ of the parameter  $\alpha$. Recall that $\Phi(k)=\max_{i\in\mathcal{V}}x_i(k) - \min_{i\in\mathcal{V}}x_i(k) $ is the consensus measure.  The trajectory corresponding to each value of $\alpha$ is shown in Figure \ref{traj}.  We see that finite-time convergence is achieved only when $\alpha$ is zero, i.e., only for the maximizing case of  Algorithm (\ref{9}).

\begin{figure}[H]
\centerline{\epsfig{figure=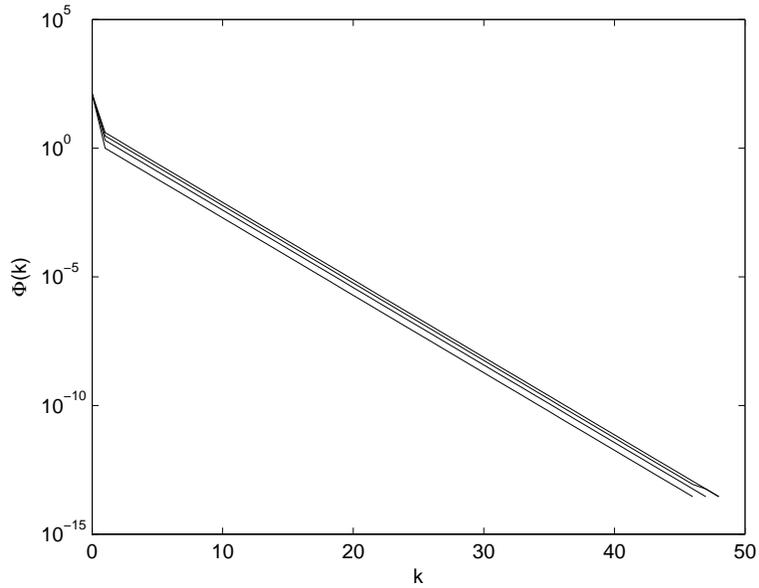, width=0.7\linewidth=0.25}}
\caption{The consensus measure $\Phi$  for nearest-neighbor graph with $\mu=123,\dots,126$  (upper curve to lower curve).  Finite-time convergence does not hold for $\mu<127$ according to Theorem~\ref{thm8}.}\label{neighborgraph}
\end{figure}

\begin{figure}[H]
\centerline{\epsfig{figure=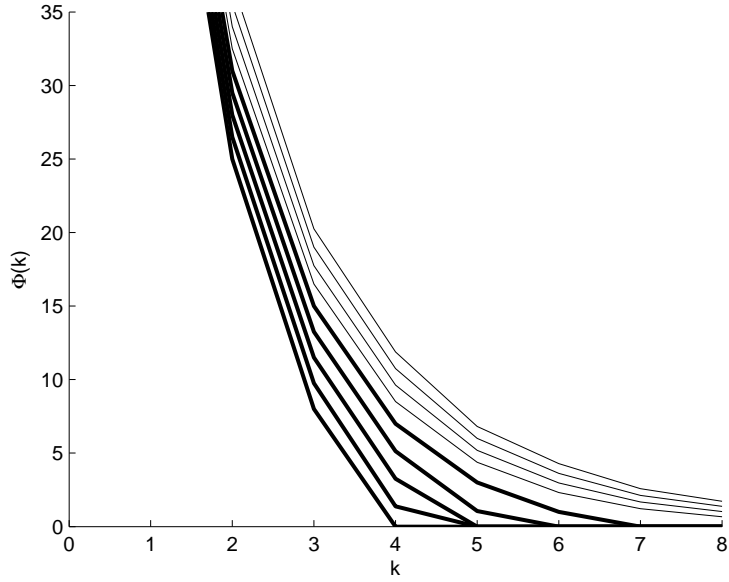, width=0.7\linewidth=0.25}}
\caption{The consensus measure $\Phi$  for nearest-value graph with $\mu=60,\dots,68$  (upper curve to lower curve).  The bold curves ($\mu=64,\dots,68$) indicate the cases with finite-time convergence. Finite-time convergence does not hold for  $\mu<64$. The simulations illustrate Theorem~\ref{thm10}.}\label{valuegraph}
\end{figure}

\noindent {\bf Example 2.} Consider a network with $n=128$ nodes. Take initial value for the $i$'th node as $x_i(0)=i$. Let $\eta_k\equiv 0$ and $\alpha_k\equiv0.5$ in  Algorithm (\ref{9}). The communication graph is given by the nearest-neighbor or nearest-value rule. The evolution of the convergence properties depending on $\mu$ is shown in Figure \ref{neighborgraph} and \ref{valuegraph}. In Figure \ref{neighborgraph}, we illustrate the conclusion in Theorem \ref{thm8}  about that finite-time convergence does not hold under nearest-neighbor graphs $\mathcal{G}^{\mu }_{x(k)}$ for $\mu<127$. In Figure \ref{valuegraph}, we illustrate the conclusion in Theorem \ref{thm10}  about that finite-time convergence does not hold under nearest-value graphs $\mathcal{G}^{\mu v}_{x(k)}$ for $\mu<64$.

\begin{figure}[H]
\centerline{\epsfig{figure=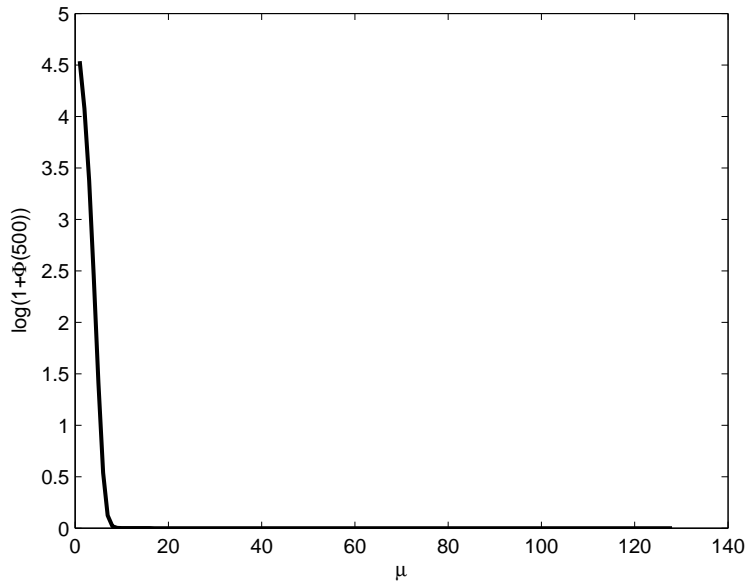, width=0.7\linewidth=0.25}}
\caption{ The convergence speed is heavily influenced by $\mu$ for nearest-value graphs as illustrated in the plot by  the value of $\Phi(k)$ after $k=500$ steps. There is a sharp increase of the convergence rate when $\mu$ increases above $5$.}\label{rate}
\end{figure}

In Figure \ref{rate}, we show the dependence  of the convergence rates on the number of neighbors for nearest-value graphs. We see that there is a sharp increase of the convergence rate when $\mu$ is larger than about five. This is consistent with the convergence speed for consensus algorithm on Cayley graphs, where the convergence rate increases very fast as the graph degree increases \cite{speranzon}.

\section{Conclusions}
This paper focused on a uniform model for distributed averaging and maximizing.  Each node iteratively updated its  state as a weighted average of its own state,  the minimal state,  and maximal  state  among its neighbors. We proved that finite-time consensus is  almost impossible for averaging   under the uniform model.  The communication graph was time-dependent or state-dependent. Necessary and sufficient conditions were established on the graph to ensure a global consensus. For time-dependent graphs, we showed that quasi-strong connectivity is critical for  averaging algorithms, as is strong connectivity for  maximizing algorithms. For state-dependent graphs defined by a $\mu$-nearest-neighbor rule, where each node interacts with  its $\mu$ nearest smaller neighbors  and the  $\mu$ nearest larger neighbors, we showed that  $\mu+1$ is a critical  number of  nodes when consensus transits from finite time to asymptotic convergence in the absence of  node self-confidence: finite-time consensus disappears suddenly when the number of nodes is larger than $\mu+1$. This critical number of nodes turned  out to be  $2\mu$ if each node chooses to connect to nodes with different values.  The results revealed the fundamental connection and difference between distributed averaging and maximizing, but  more challenges still lie in the  principles underlying the two types of algorithms, such as their convergence rates.

\vspace{10mm}

\begin{center}
{\Large Acknowledgment }
\end{center}
The authors would like to thank Dr. Kin Cheong Sou for his generous help on the numerical examples.

\end{document}